# Trigonal layered rosiaite-related antiferromagnet MnSnTeO$_6$: ion-exchange preparation, structure and magnetic properties

V.B. Nalbandyan, M.A. Evstigneeva, T.M. Vasilchikova, K.Yu. Bukhteev, A.N. Vasiliev and E.A. Zvereva

Ion-exchange treatment of Na$_2$SnTeO$_6$ in molten salt mixtures resulted in rosiaite (PbSb$_2$O$_6$)-related MnSnTeO$_6$. Its crystal structure was refined by the Rietveld method. Of the three possible models of Sn/Te distribution, the disordered model (($P\bar{3}1m$, $a$ = 5.23093(11) Å, $c$ = 4.62430(16) Å) was preferred based on bond distances. However, it is supposed that each individual (SnTeO$_6$)$^{2-}$ layer retains complete ordering of the precursor and the apparent disorder is only due to stacking faults. The magnetic studies have shown that MnSnTeO$_6$ orders antiferromagnetically at Néel temperature $T_N$ = 9.8 K. The effective magnetic moment reasonably agrees with theoretical estimations assuming high-spin configuration of Mn$^{2+}$ ($S$ = 5/2). Electron spin resonance reveals spin dynamics in accordance with aniferromagnetic ordering with critical broadening of ESR linewidth indicating the low-dimentional type of exchange interactions. Based on the temperature and field-dependent magnetization studies, the magnetic phase diagram of the new compound was constructed.

## 1. Introduction

The layered oxides of magnetic transition metals (quasi two dimensional (2D) magnetic systems) with triangular spin lattices are currently a research topic of tremendous interest due both to the unique physics and to their potential in device applications.[1-5] Intensive theoretical and experimental efforts have been applied to understand such frustrated spin systems, which along with the honeycomb spin systems with strong spin-orbit coupling can provide the realization of the elusive spin liquid (SL) state.[6-9] SLs evade conventional long-range order or symmetry breaking down to $T \to 0$.[6] These highly entangled states have unique properties, including possible topological order or fractional excitations (such as spinons and Majorana fermions). Nowadays promising spin liquid candidates are supposed to be the antiferromagnets on the edge-sharing triangular lattice. A commitment to rigorous geometric landmarks for new triangular lattice frustrated materials is important, that remains a challenge task in synthesis. Along with disordered liquids the triangular lattice layered magnets may order into exotic long-range ordered states, which are stabilized by interlayer interactions. The real quantum ground state of novel triangular spin systems are hardly predictable since it strongly depends on competition of the frustration and anisotropy.

In search for new magnetic phases with triangular lattice, trigonal layered structure type of rosiaite (PbSb$_2$O$_6$) attracted our attention. Previously known magnetic members of this structural family are M$^{2+}$As$_2$O$_6$ (M = Mn, Co, Ni, Pd).[10-12] In the MSb$_2$O$_6$ series, with larger Sb(5+) substituting for small As(5+), M$^{2+}$ cation should also be larger, and all stable MSb$_2$O$_6$ rosiaites (M = Cd, Ca, Sr, Pb, and Ba) are diamagnetic. However, smaller magnetic cations might be introduced via low-temperature ion exchange in ilmenite-type NaSbO$_3$, and metastable rosiaite-type MSb$_2$O$_6$ (M = Mn, Co, Ni, and Cu) were prepared and studied.[13,14]

In this work, we apply the same approach to ion exchange in Na$_2$SnTeO$_6$.[15] It is structurally related to but not strictly isostructural with ilmenite NaSbO$_3$. Both are based on honeycomb-type anionic layers of edge-sharing SbO$_6$, SnO$_6$ and TeO$_6$ octahedra. Sodium ions reside between these layers, also in distorted octahedra, sharing one face with SbO$_6$ or, respectively, TeO$_6$ octahedron. The two structures differ in two aspects:

(i) NaSbO$_3$ is rhombohedral triple-layered whereas Na$_2$SnTeO$_6$ is double-layered, space group $P\bar{3}1c$;
(ii) Sn and Te are completely ordered on Sb positions.

Nevertheless, we suggested that structural transformation to the rosiaite-related structure may be accomplished by layer gliding during divalent ion exchange in Na$_2$SnTeO$_6$ like in NaSbO$_3$, and this was confirmed experimentally.

## 2. Experimental

### 2.1. Sample preparation and elemental analysis

All chemicals used were of reagent-grade. MnSO$_4$·xH$_2$O, SnO$_2$ and TeO$_2$ were calcined in air for 2 h at 400 °C, and Mn$_2$O$_3$, at 750 °C. Hydrous MnCl$_2$ was dried in vacuum at 200 °C. Na$_2$CO$_3$, KCl, and CsCl were dried in air at 150 °C. All dehydrated materials were stored in closed containers in a desiccator. Na$_2$SnTeO$_6$ was prepared by solid-state reactions according to slightly modified literature method.[15] First of all, Na$_2$SnO$_3$ was synthesized from stoichiometric quantities of Na$_2$CO$_3$ and SnO$_2$ by solid-state reactions at 800 and 950 °C with intermediate grinding and pressing until peaks from SnO$_2$ were completely eliminated from the X-ray diffraction (XRD) powder pattern. We avoided temperatures above 1000 °C that led to sodium deficiency in the reported final product.[15] Then, a stoichiometric amount of TeO$_2$ was carefully admixed, the mixture pressed into thin discs to facilitate oxidation by air and calcined at 750 °C twice for 2 h with intermediate regrinding and pressing. The XRD pattern of thus obtained Na$_2$SnTeO$_6$ (Fig. 1a) was identical to the literature data and refined to the lattice constants ($a$ = 5.3416(1), $c$ = 10.6991(3) Å) in good agreement with the reported values.

The powdered Na$_2$SnTeO$_6$ was then ion-exchanged with alow-melting salt mixture based on Mn$^{2+}$,K$^+$||SO$_4^{2-}$,Cl$^-$ reciprocal system, as in the preceding work,[13] with Mn$^{2+}$ taken in excess against the stoichiometry. The mixtures were held at 400-450 °C for 40-80 min in Ar atmosphere. After cooling, all salts were dissolved in warm water, the precipitate washed by decantation and dried at 150 °C.

For elemental analysis, an electron microprobe (INCA ENERGY 450/XT) was used with an X-Act ADD detector based on an electron microscope VEGA II LMU (Tescan) operated at the accelerating voltage of 20 kV. Due to considerable scatter, the molar ratios Mn/Te, Sn/Te, and Na/Te were averaged on 14 measurements at various points of each sample. No potassium was detected.

## 2.2. Diffraction studies

XRD measurements were done using an ARL X'TRA diffractometer equipped with a solid-state Si(Li) detector. Lattice parameters were refined with CELREF 3 (J. Laugier & B. Bochu) after angular corrections with an internal standard, corundum NIST SRM 676. For structural refinements by the Rietveld method, the GSAS + EXPGUI suite[16,17] was used. To reduce grain orientation, the sample was mixed with amorphous powder (instant coffee).

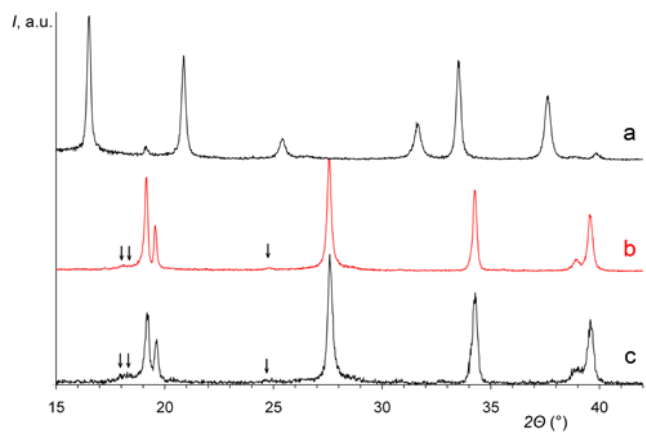

**Fig. 1.** Low-angle part of the XRD patterns of starting $Na_2SnTeO_6$ (a) and two ion-exchange products, 1 and 2 (b, c)

**Table 1.** Comparison of the XRD and EDX data for the two $MnSnTeO_6$ samples studied in the present work. Standard uncertainties of the last decimal digit are given in parentheses.

| Sample | $a$, Å | $c$, Å | Mn/Te | Sn/Te | Na/Te |
|---|---|---|---|---|---|
| 1 | 5.2283(8) | 4.6169(8) | 0.96(17) | 1.08(18) | 0.14(6) |
| 2 | 5.2313(20) | 4.6109(9) | 0.90(3) | 1.01(2) | 0.15(5) |

## 2.3. Magnetic measurements

The magnetic measurements (temperature dependencies of the magnetic susceptibility and magnetization isotherms) were performed by means of a Quantum Design PPMS 9 system.

The temperature dependences of magnetic susceptibility were measured at the magnetic field $B$ = 0.1 – 9 T in the temperature range 2 – 300 K. The isothermal magnetization curves were obtained for magnetic fields $B \leq 9$ T at $T$ = 2 – 10 K after cooling the sample in zero field.

Electron spin resonance studies were carried out using X-band ESR spectrometer CMS 8400 (ADANI) (f ≈ 9.4 GHz, B ≤ 0.7 T) equipped with a low-temperature mount, operating in the range $T$ = 5 – 500 K. The effective g-factor has been calculated with respect to BDPA (a,g-bisdiphenyline-b-phenylallyl) reference sample with $g_{et}$ = 2.00359.

## 3. Results and discussion

### 3.1. Sample preparation and identification

To avoid side reactions during ion-exchange in molten salts (e.g., incongruent dissolution leading to formation of $SnO_2$), it was important to reduce both duration and temperature of the process. In absence of detailed phase diagrams for the multicomponent reciprocal salt systems, various additions to the basic $MnSO_4$-KCl mixture[13] were tested to reduce its melting point. So far, the most phase-pure $MnSnTeO_6$ samples were obtained from the mixture $Na_2SnTeO_6$ + 6$MnSO_4$ + 6KCl + 6CsCl for 1 h at 430 °C. The two samples prepared in these nominally identical conditions contained only a minor amount of unidentified foreign phase (Fig. 1 b,c) and showed similar XRD and EDX results (Table 1). Sample 1 was used for structural analysis and sample 2, for magnetic studies.

Deviation of Mn/Te ratio from unity seems to be within experimental accuracy, but considerable sodium impurity requires explanation. By analogy with $MnSb_2O_6$ prepared previously in a similar way[13], we reject Na substitution for Mn because electroneutrality requires two $Na^+$ for one $Mn^{2+}$ and the crystal structure has no place for the extra cations. More likely, the sodium impurity is present in residual $Na_2SnTeO_6$ layers existing as stacking faults in the main phase, in occluded salts, and/or in the foreign phase indicated in Fig. 1 b,c.

## 3.2. Crystal structure

Although Sn and Te have very similar scattering factors for both X-rays and neutrons, their ordering in the starting $Na_2SnTeO_6$ is proved definitely by strong difference in bond lengths (0.133 Å) and also by $^{125}$Te MAS NMR.[15] Therefore, it seemed natural that this ordering would persist during the low-temperature $Mn^{2+}$ ion exchange. There are three possible models for arrangement of the two cations substituting for Sb in $PbSb_2O_6$ (rosiaite) structure type shown in Fig. 2:
(a) alternation of Sn and Te along the three-fold axis giving rise to a superlattice with doubled $c$, as in $LaFeTeO_6$, space group $P\bar{3}1c$;[18] (b) arrangement of Sn under Sn and Te under Te, as in the $KNiIO_6$ type, space group $P312$;[19]
(c) disordered stacking, i.e., random intergrowth of structures (a) and (b), that should be described as rosiaite aristotype, $P\bar{3}1m$, with mixed $Sn_{0.5}Te_{0.5}$ occupancy on Sb sites.

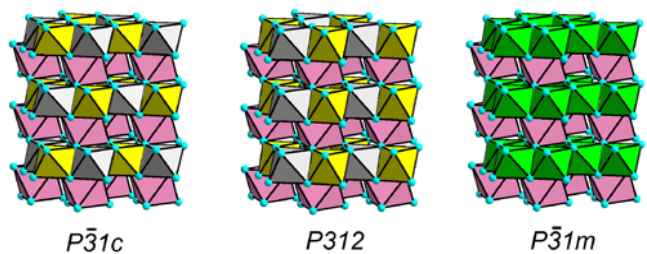

**Fig. 2.** Three structural models for $MnSnTeO_6$. Cyan spheres, oxygen; pink octahedra, $MnO_6$; grey octahedra, $SnO_6$; yellow octahedra, $TeO_6$; green octahedra, mixed $(Mn,Sn)O_6$.

**Table 2**. Comparison of the three structural models for rosiaite-type $MnSnTeO_6$

| Model | Average bond lengths, Å | | | | Discrepancy factors | | |
|---|---|---|---|---|---|---|---|
| | Te-O | Sn-O | (Te,Sn)-O | Mn-O | $R_p$ | $R_{wp}$ | $\chi^2$ |
| (a) $P\bar{3}1c$ | 1.955 | 1.993 | - | 2.288 | 0.140 | 0.187 | 4.92 |
| (b) $P312$ | 1.946 | 2.012 | - | 2.280 | 0.146 | 0.189 | 4.915 |
| (c) $P\bar{3}1m$ | - | - | 1.999 | 2.237 | 0.122 | 0.166 | 3.826 |
| Sum of ionic radii[20] | 1.92 | 2.05 | 1.985 | 2.19 | | | |

Due to similarity of scattering factors for Sn and Te, no considerable superlattice reflections can be expected in model (a), whereas models (b) and (c) do not generate superlattice reflections at all. Therefore, discrimination between the three models can only be done by careful refinement of oxygen coordinates in each model followed by comparison based on bond lengths and agreement factors.
The results listed in Table 2 show that both ordered models provide only small bond length differences (0.038 and 0.066 Å) compared to the expected value of 0.13 Å (observed in $Na_2SnTeO_6$[15] experimentally). In addition, these models result in elevated discrepancy factors. Thus, complete ordering is not supported and the disordered model (c) is selected. The profile refinement results are shown in Fig. 3. Further details may be found in Electronic supplementary information (cif).

Despite selection of the model (c), we believe that each individual layer remains completely ordered and the Sn/Te mixing is only fictitious due to stacking faults generated by layer gliding during ion exchange. Woodward et al.[15] noticed that $NaO_6$ octahedra in isostructural $Na_2MTeO_6$ phases share one face with the smaller of other two cations: Ge(4+) in $Na_2GeTeO_6$ and Te(6+) in $Na_2SnTeO_6$. This is one of the factors maintaining ordered stacking sequence in Na$_2$MTeO$_6$. Going to MnSnTeO$_6$, face sharing vanishes and stacking disorder becomes more probable.

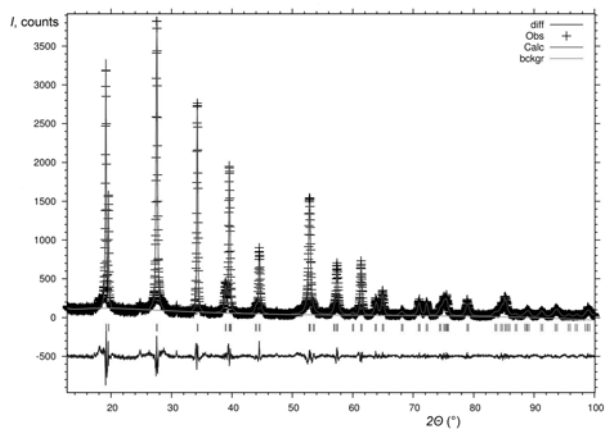

**Fig. 3.** Results of the XRD Rietveld refinement for MnSnTeO$_6$. $P\bar{3}1m$, $a$ = 5.23064(11), $c$ = 4.62476(18). Crosses, experimental data; solid line, calculated profile; line in the bottom, difference profile; vertical bars, positions of Bragg reflections.

### 3.3. Static magnetic properties

The static magnetic susceptibility $\chi = M/B$ of MnSnTeO$_6$ is presented in Fig. 4. The pronounced maximum at $T_N$ = 9.8 K indicates the onset of long-ranged antiferromagnetic (AFM) order.

Suggesting Curie–Weiss-type behaviour in the paramagnetic phase, we have analysed the high temperature part of $\chi$ (T) according to the Curie–Weiss law with addition of a temperature-independent term $\chi_0$:

$$\chi = \chi_0 + \frac{C}{T-\Theta} \quad (1)$$

where $C$ is the Curie constant and $\Theta$ is the paramagnetic Curie–Weiss temperature. The negative Weiss temperature $\Theta$ = -20.7 ± 1 K indicates the dominant antiferromagnetic interaction in this compound.

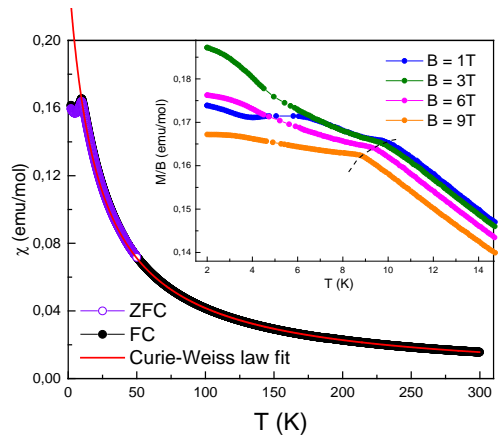

**Fig. 4.** Temperature dependence of magnetic susceptibility $\chi = M/B$ at $B$ = 0.1 T for MnSnTeO$_6$ recorded in zero-field-cooled ZFC (filled symbols) and field-cooled FC (open symbols). The solid red curve is approximation in accordance with the Curie-Weiss law (eqn. 1). Inset: $M/B$ (T) curves for MnSnTeO$_6$ at various external magnetic fields.

The effective magnetic moment $\mu_{eff}$ can be estimated from the Curie constant in accordance with:

$$\mu_{eff}^2 = 3k_B C/\mu_B^2, \quad (2)$$

where $k_B$, $\mu_B$ and $N_A$ are the Boltzmann constant, the Bohr magneton and the Avogadro number, respectively. To reduce the number of variable parameters, the diamagnetic contribution in MnSnTeO$_6$ was estimated independently as $\chi_0 \approx$ -1.06×10$^{-4}$ emu/mol by summing Pascal's constants[21] taken for each individual atom constituting MnSnTeO$_6$. In this case, the resulting value of $\mu_{eff}$ = 6.05 $\mu_B$/f.u. slightly exceeds the theoretical estimate $\mu_{theor}$ = 5.92 $\mu_B$/f.u., suggesting that magnetism in MnSnTeO$_6$ is associated with Mn$^{2+}$ ions ($S$ = 5/2):

$$\mu_{theor}^2 = g^2 \mu_B^2 n S(S+1), \quad (3)$$

where $g$ = 2.00 ± 0.01 is the g-factor (experimentally determined in the present work from the ESR data – see Section 3.4). When set all parameter in eqn.(1) to be variable, the effective magnetic moment $\mu_{eff}$ calculated from corresponding Curie constant was found to be in better agreement with theoretical estimates.

The $M/B$ (T) dependences in the low temperature range under variation of the magnetic fields up to 9 T are shown in the inset of Fig. 4. With increasing the magnetic field, the maximum at $T_N$ is markedly rounded and shifts toward the low-temperature side.

As can be seen from Fig. 5, the magnetization isotherms $M$ ($B$) do not display hysteresis or saturation in magnetic fields up to 9 T. Within this range of the applied magnetic fields, the magnetic moment is still far below the theoretically expected saturation magnetic moment for Mn$^{2+}$ ($S$ = 5/2):

$$M_S = gS\mu_B \approx 5\mu_B \quad (4)$$

However, the magnetization curves have a slight upward curvature suggesting the possible presence of a magnetic field induced spin-flop transition which is usually associated to magnetic anisotropy characteristic to easy-axis antiferromagnets. Also this transition is manifested in *M*/*B* (*T*) curve at *B* = 1 T. The critical field has been determined from maximum on derivative dependence d*M*/d*B* (*B*) and amounts to $B_{SF}$ = 1.16 T. It weakly decreases with increasing temperature and eventually disappears at $T > T_N$.

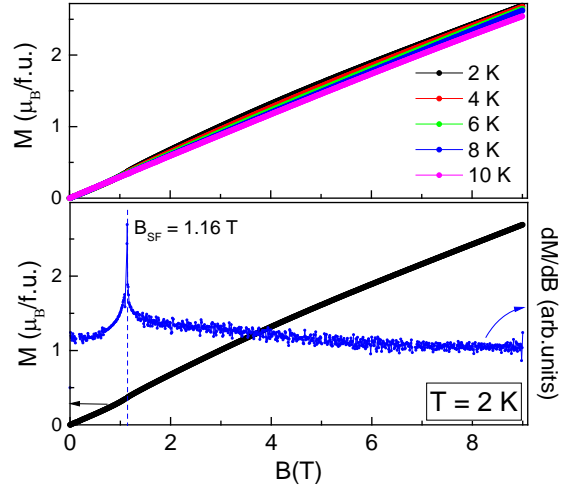

**Fig. 5.** Magnetization curves for MnSnTeO$_6$ at various temperatures (upper panel). The *M* (*B*) isotherm at *T* = 2 K and corresponding magnetization derivative d*M*/d*B* shown at lower panel.

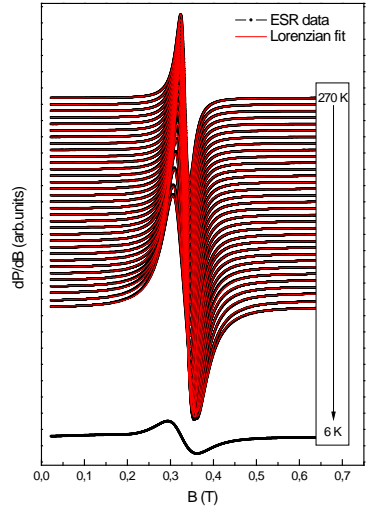

**Fig. 6.** Evolution of the ESR spectra for MnSnTeO$_6$ with decreasing temperature. The symbols represent experimental data, solid red curves are approximations of the ESR spectrum using Lorenzian profile.

### 3.4. Dynamic magnetic properties

Evolution of the ESR spectra of powder sample MnSnTeO$_6$ with temperature is presented in Fig. 6. These spectra in the paramagnetic phase ($T > T_N$) present a single Lorentzian shape line apparently ascribable to Mn$^{2+}$ ions. The main ESR parameters were deduced by fitting the experimental spectra with Lorentzian profile taking into consideration two circular components of the exciting linearly polarized microwave field on both sides of *B* = 0 since the line observed is relatively broad:[22]

$$\frac{dP}{dB} \propto \frac{d}{dB}\left[\frac{\Delta B}{\Delta B^2+(B-B_r)^2} + \frac{\Delta B}{\Delta B^2+(B+B_r)^2}\right] \quad (5)$$

This is a symmetric Lorenzian line, where *P* is the power absorbed in the ESR experiment, *B* – magnetic field, $B_r$ – resonance field, Δ*B* – the linewidth. Results of the ESR lineshape fitting are shown by red solid lines in Fig. 6. Evidently, the fitted curves are in good agreement with the experimental data. Effective g-factor *g* = 2.00 ± 0.1 was found to be isotropic and to remain almost constant with decreasing temperature down to ~ 70 K (upper part of Fig. 7). Upon further cooling, the shift of the resonance field occurs indicating appearance of the short-range fluctuations on approaching long-range ordering transition from above. In contrast, the linewidth $\Delta B$ demonstrates continuous increase with decreasing the temperature (lower part of Fig. 7). This behaviour implies the presence of extended range of strong short-range correlations essentially higher than $T_N$, those are characteristic of the frustrated and low-dimension systems.

To complete the analysis of the ESR spectra, we have also estimated the integral ESR intensity, which is known to be proportional to the concentration of paramagnetic centres, by double integration of the first derivative of the absorption line for each temperature. It was found that the temperature dependence of the normalized integral ESR intensity $\chi_{ESR}(T)$ agrees quite well with the static magnetic susceptibility data $\chi(T)$.

### 3.5. Critical behaviour of ESR linewidth

Upon cooling, the linewidth $\Delta B$ continuously increases (lower panel on Fig. 7) which may be due to critical broadening associated with the evolution of spin-spin correlations and concomitant slowing down spin fluctuations upon approaching the Neel temperature from above. Such line broadening can be described in terms of a power law of the reduced temperature, with the critical exponent $\beta$, which reflects to the dimensionality of the magnetic subsystem.

In this case, the temperature variation of $\Delta B$ can be described by[23-26]

$$\Delta B(T) = \Delta B^* + A \cdot \left[\frac{T_N^{ESR}}{T-T_N^{ESR}}\right]^{\beta} \quad (6)$$

where the first term $\Delta B^*$ describes the high-temperature exchange narrowed linewidth, which is temperature independent, while the second reflects the critical behaviour with $T^{ESR}_N$ being the temperature of the order-disorder transition. As one can see from the log plot (Fig. 7) the fitting of the experimental data over the whole temperature range higher the Neél temperature (blue dashed curve) implies
failure of a power-law description both in the high temperature limit and upon approaching $T_N$ at $T < 50$ K.

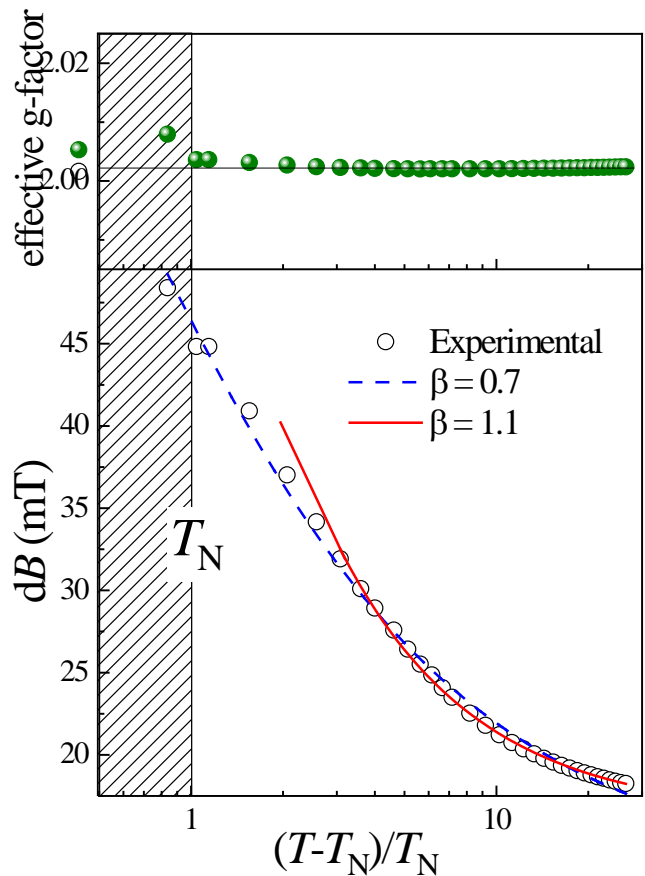

**Fig. 7.** Temperature dependencies of the effective g-factor and the ESR linewidth $\Delta B$ for MnSbTeO$_6$ derived from the ESR data. Circles are experimental data, solid and dashed lines are approximation in accordance of power law (see Eq. 6) as described in the text.

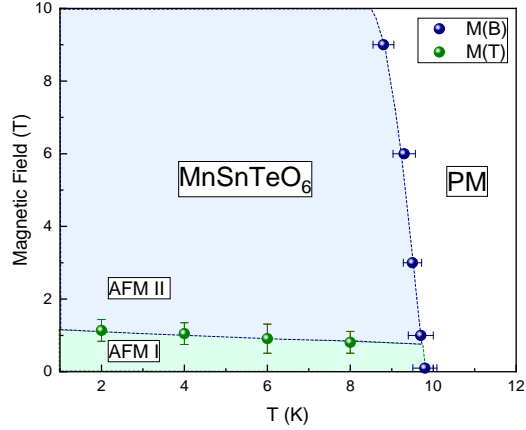

**Fig. 8.** The magnetic phase diagram for new trigonal layeredtellurate MnSnTeO$_6$

At the same time the approximation using the range $T > 50$ K describes the data reasonably well with $\beta = 1.1(1)$. The critical divergence becomes weaker upon approaching the 3D spin ordered phase. The observed critical exponent is three times larger than what is theoretically expected for 3D antiferromagnets[23,24] and is close to values for 2D oxides[1] like for example experimentally observed for other quasi 2D layered tellurates and antimonates Li$_4$NiTeO$_6$ ($\beta = 1.59$)[27], Li$_3$Ni$_2$SbO$_6$ ($\beta = 0.9$)[28] or Na$_4$FeSbO$_6$ ($\beta = 1.01$)[29].

### 3.6. Magnetic phase diagram

Based on the data of static and dynamic magnetic studies performed in this paper, a magnetic phase diagram was constructed for MnSnTeO$_6$ (Fig. 8). In zero magnetic field, the paramagnetic phase is realized at temperatures higher than 10 K. Applying the magnetic field slightly shifts the phase boundary corresponding to this transition to lower temperatures. It is also evident from the $B - T$ diagram, that the ground state of MnSnTeO$_6$ is found to be antiferromagnetic but an intermediate magnetic phase (II) was induced by applying a magnetic field. Thus, there are at least two phases (I and II) related probably to two different spin orientations. It can be associated with the change of the mutual orientation of neighbouring spins on the triangular lattice. Note, that the critical magnetic field takes relatively small value implying moderate anisotropy. When the magnetic field overcomes the energy of anisotropy the spin-rotation situation occurs and one of the spin configurations (AFM I at B = 0) is destabilized, i.e., it appears only in the case when the anisotropy is dominant in comparison with external field.

## 4. Conclusion

Using ion exchange between Na$_2$SnTeO$_6$ and a low-melting salt mixture, we have prepared a new layered manganese tin tellurate MnSnTeO$_6$. Its trigonal structure is analogous to that of rosiaite PbSb$_2$O$_6$. Although XRD profile refinement could not reveal Sn/Te ordering on Sb sites, it is assumed that each individual (SnTeO$_6$)$^{2-}$ layer is completely ordered and apparent Sn/Te disorder is due to stacking faults.

The static magnetic susceptibility data indicate the onset of AFM order at $T_N \approx 9.8$ K. The high-temperature magnetic susceptibility data exhibit the Curie–Weiss behaviour with a Weiss temperature taking a negative value $\Theta = -20.7 \pm 1$ K that indicates a predominance of the antiferromagnetic coupling. Upward curvature of the magnetic isotherms of $M(B)$ with the temperature variation indicates the presence of a spin-flop transition at $B_{SF} = 1.16$ T induced by the magnetic field. Structural disorder due to stacking faults does not affect the magnetic properties. In particular, the perfect coincidence of ZFC and FC magnetic susceptibility confirms the absence of any cluster effects.

ESR spectra in the paramagnetic phase show a single Lorenzian shape line attributed to Mn$^{2+}$ ions in octahedral oxygen coordination characterized by the isotropic effective g-factor $g = 2.00 \pm 0.1$. The temperature dependence of the ESR integral intensity, the line width, and the shift of the resonant field point out to the extended range of the short-range correlations in a wide *T*-range higher long-range order. Increase in the ESR linewidth with temperature decreasing is in agreement with typical for AFM compounds critical broadening and the critical exponent corresponds rather 2D character of AFM magnetic correlations.

Summarizing the results of magnetic studies, a magnetic phase diagram of the new 2D triangular magnet MnSnTeO$_6$ has been constructed. The presence of two different ordered AFM phases at low temperatures is in accordance with anisotropic character of the magnetic coupling for this new compound.

## Acknowledgements

The work was supported by the Russian Foundation for Basic Research through the grant 18-03-00714 and also by the Russian Ministry of Education and Science of the Russian Federation through NUST «MISiS» grant К2-2017-084 and by the Act 211 of the Government of Russia, contracts 02. A03.21.0004 and 02.A03.21.0011. E.A.Z. acknowledges Russian Science Foundation through the project 17-12-01207 for support of the static and dynamic magnetic studies. The authors thank Drs S.I. Shevtsova and Yu.V. Popov for the EDX analysis.